\documentclass{aa}
\usepackage{times,amssymb,amsmath}
\usepackage{epsfig}

\title{Probing the faint end of the Galaxy luminosity function at z=3
with Ly$\alpha$ emission
\thanks{Based on observations made with ESO Telescopes at the  
Paranal Observatory under programme ID 64.O-0187}
}
\titlerunning{The faint end of the Galaxy luminosity function at z=3}
\author{J.U. Fynbo \inst{1}
   \and P. M\o ller \inst{1}
   \and B. Thomsen \inst{2}
}

\institute{
           European Southern Observatory,
           Karl-Schwarzschild-Stra\ss e 2,
           D-85748 Garching, Germany
           \and
           Institute of Physics and Astronomy,
           \AA rhus University, DK-8000 \AA rhus C., Denmark
           }
\offprints{J. U. Fynbo}
\mail{jfynbo@eso.org}

\date{Received  / Accepted }

\abstract{
We present spectroscopic observations obtained with the ESO 
Very Large Telecope (VLT) of seven candidate Ly$\alpha$ emitting 
galaxies in the field of the radio quiet Q1205-30 at z=3.04 
previously detected with deep narrow
band imaging. Based on equivalent widths and limits on line
ratios we confirm that all seven objects are Ly$\alpha$ emitting
galaxies.
Deep images also obtained with the VLT in the B and I bands 
show that five of the seven galaxies have very faint continuum fluxes
(I(AB)$\approx$26.8 and B(AB)$\approx$27.3). 
The star formation rates of these seven galaxies estimated from the
rest-frame UV continuum around 2000\AA, as probed by the I-band 
detections, as well as from the Ly$\alpha$
luminosities, are 1--4 M$_{\sun}$ yr$^{-1}$ assuming a 
Hubble constant of 65 km s$^{-1}$ Mpc$^{-1}$, $\Omega_m=0.3$, and 
$\Omega_{\Lambda}=0.7$. This is 1--3 orders of magnitude lower 
than for other known populations of star-forming galaxies at similar 
redshifts (the Lyman-Break galaxies and the sub-mm selected sources). 
The inferred density of the objects is high,
16$\pm$4 per arcmin$^2$ per unit redshift. This is consistent with
the integrated luminosity function for Lyman-Break galaxies 
down to R=27 if the fraction of Ly$\alpha$ emitting galaxies
is $\approx$70\% at the faint end of the luminosity function. However, 
if this fraction is 20\% as reported for
the bright end of the luminosity function then the space density
in this field is significantly larger (by a factor of 3.5) than
expected from the luminosity function for Lyman-Break galaxies in
the HDF--North. This would be an indication that at least some 
radio quiet QSOs at high redshift reside in overdense environments
or that the faint end slope of the high redshift 
luminosity function has been underestimated.
We find evidence that the faint Ly$\alpha$ galaxies
are essentially dust--free.
These observations show that Ly$\alpha$ emission is an efficient
method by which to probe the faint end of the luminosity function
at high redshifts.
}

\begin{document}

\maketitle
\section{Introduction}
During the last ten years the amount of observational data neccessary
for understanding galaxy formation at high redshift has increased
from almost nothing to a stage where of order a thousand galaxies has
been detected in emission and confirmed spectroscopically. The bulk
of these galaxies has been selected by use of the Lyman-break colour
selection technique (Steidel and Hamilton 1992; Steidel et al. 1996;
Adelberger and Steidel 2000) and later been spectroscopically confirmed.
For the Lyman-break galaxies (LBGs) it is important to keep in mind
that only those brighter than the ``spectroscopic limit''
(R(AB)$\lesssim$25.5) may have their redshifts confirmed, unless a
Ly$\alpha$ emission line is present. The spectroscopically confirmed
LBGs (SLBGs) have star formation rates similar to local starburst
galaxies. The brightest end of the high redshift LBG luminosity
function (LF) is therefore well determined, while the faint end is
based on photometric redshifts and significant corrections for
incompleteness.
One may improve the photometric redshift selection by inclusion of
infrared colours (e.g. Fontana et al. 2000 and references therein).
A second selection technique for high redshift galaxies is selection
by means of sub-mm emission. This technique is at present only
sensitive to galaxies with star-formation rates exceeding
10$^3$ M$_{\sun}$ yr$^{-1}$ (e.g. Ivison et al. 2000).
Independent of their emission properties, objects found at high
redshifts as Damped Lyman-$\alpha$ Absorbers (DLAs) in the spectra of
background QSOs are likely related to galaxies or to galaxies in the
process of assembly. DLAs are in current models of galaxy formation
found to trace the collapsed regions near intersections of dark-matter
filaments where galaxies form (e.g. Rauch et al. 1997). Furthermore,
DLAs are chemically enriched implying that star-formation is or has
been going on nearby. 

Several studies have shown that Ly$\alpha$ narrow-band imaging is an
alternative technique to identify high redshift galaxies
(M\o ller \& Warren 1993; Francis et al. 1995; Pascarelle et al. 1996;
Pascarelle et al. 1998; Cowie and Hu 1998; Hu et al. 1998; Fynbo et al. 
1999, 2000; Kudritzki et al. 2000; Kurk et al. 2000; Pentericci et al. 
2000, Steidel et al. 2000; Roche et al. 2000). So far it has not been 
considered efficient enough to seriously compete with the LBG technique, 
but it does have the advantage that spectroscopic confirmation is not 
limited by the broad band flux.
Most recently the current success in identification of Gamma-Ray
Bursts at high redshifts, has provided a completely independent
selection technique for the host galaxies of Gamma-Ray Bursts
(e.g. Odewahn et 
al. 1998; Bloom et al. 1999; Holland and Hjorth 1999; Vreeswijk et al. 
2000; Smette et al. 2001; Jensen et al. 2001; Fynbo et al.  2001). 


From the point of view of observations, what is still missing is an 
understanding of the connection between high redshift galaxies selected 
in different ways. Fontana et al. (2000) find that photometric redshifts
including IR colours select nearly a factor of 2 more high redshift
galaxy candidates in the Hubble Deep Fields (HDFs) than pure LBG
photometric selection does, to the same flux limits. However, this
discrepancy cannot be resolved spectroscopically. While LBGs are
selected by continuum flux, DLAs are selected by gas cross-section.
Under the assumption 
of a scaling relation between the gas disc size and the luminosity for 
high redshift galaxies, and by normalising this relation using the few
observed impact parameters for z$\approx$3 DLAs, DLAs are predicted to 
be much fainter than the SLBG limit
(Fynbo et al. 1999; Haehnelt et al. 2000; see also Ellison et al. 2001). 
Members of the population of galaxies producing 
DLAs are obviously not only found close to QSO lines of sight so we
expect an abundant population of galaxies below the current
spectroscopic limit for LBGs. 

From a theoretical point of view, the 
properties of this faint end of the high redshift LF
is important in order to constrain the importance of 
stellar feedback processes (e.g. Efstathiou 2000; Thacker and Couchman 
2000; Poli et al. 2001) 
and the fraction of the background of hydrogen ionising photons 
produced by high mass stars at high redshift (Steidel et al. 2001;
Haehnelt et al. 2001).

To probe the faint end of the LF at high redshifts we need methods to
search for emission from high redshift galaxies fainter than
R$\lesssim$25.5. Here several methods are possible: 
{\it i)} Using the Lyman break technique and photometric redshifts,
LFs for z=3 galaxies has been presented by 
Adelberger and Steidel (2000) and by Poli et al. (2001) down to R=27.
The faint end (R$>$25.5) of these LFs are uncertain due to the lack
of spectroscopic confirmation,
{\it ii)} Ly$\alpha$ narrow--band imaging.
{\it iii)} Imaging of DLAs at very faint continuum levels with the {\it 
Hubble Space Telescope} (M\o ller and Warren 1998;
Kulkarni et al. 2000, 2001; Ellison et al. 2001;
Warren et al. 2001; M\o ller et al. 2001 in prep), and {\it iv)} deep 
searches for the host galaxies of well-localized (to within a fraction
of an arcsec using the positions of optical transients) Gamma-Ray
Bursts. In this paper we focus on Ly$\alpha$ emission as a method by 
which to study the faint end of the LF at z=3. 

In Fynbo et al. (2000, paper I) we reported on six faint candidate 
Ly$\alpha$ Emitting Galaxies detected in a very deep narrow band image
of the z=3.036 radio quiet QSO Q1205-30 obtained with the NTT.
Here we present photometric 
and spectroscopic follow-up observations of these 6 candidates (called
S7--S12, detected at better than 5$\sigma$) plus 2 marginal candidates 
(S13 and S14, detected at the $\approx 4\sigma$ level). This paper
is concerned mostly with the details of the observations, data
reduction and the results concerning the physical properties of the
Ly$\alpha$ galaxies.
In a separate {\it Letter} we discussed the spatial distribution of the
confirmed Ly$\alpha$ galaxies and how it related to current models of
structure formation in the early universe (M\o ller and Fynbo 2001).
In that paper we concluded
that the Ly$\alpha$ emitting objects are proto-galatic sub-units in
the process of assembly, and for that reason chose to refer to them as
``Ly$\alpha$ Emitting Galaxy-building Objects'' (LEGOs). Here, for
consistency, we shall
adopt the same acronym. The rest of the paper is organized as follows:
In Sect. 2. we describe the observations and the
data reduction, in Sect. 3 we present our results, in Sect. 4 we
discuss our results in terms of star--formation rates and luminosity
function compared to that of the LBGs and in Sect. 5 we summarise our
conclusions.
Throughout this paper we adopt a Hubble constant of 65 km s$^{-1}$
Mpc$^{-1}$ and assume $\Omega_m=0.3$ and $\Omega_{\Lambda}=0.7$.

\section{Observations and data reduction}
\subsection{Preimaging and construction of masks}
For the spectroscopy we used the Multi Object Spectroscopy
(MOS) mode of FORS1 on the VLT--UT1 telescope.
Pre--imaging for MOS was obtained in Service Mode during
two nights in January 2000, well ahead of our Visitor Mode run, and
consisted of 13 exposures each of 400sec in the Bessel B-filter.
The 5$\sigma$ detection limit of the final combined pre--image was
B(AB)=26.7. Surprisingly this did not, however, allow the detection of
continuum emission from all of the eight target objects S7-S14
(see Sect. 3.2 below). 
Furthermore, for one of those objects (S9) where we {\it did}
detect continuum emission, there was a small but significant offset
between the continuum position and the narrow--band position on the
sky. For each of the eight target objects we therefore derived local
transformations between the NTT narrow--band
image and the VLT B--band image, and placed artificial stars at the
positions in the VLT image corresponding to the positions of S7-S14
in the NTT narrow--band image. Two of us did this independently and
found positions that agreed to within 1/3 of a FORS1 pixel,
corresponding to 0.07 arcsec. To minimize slitlosses, which could
become severe if our invisible objects were not perfectly centered in
the slitlets, and/or if the objects were extended in Ly$\alpha$,
we chose to use a slit--width of 1.2 arcsec for all slitlets.

We chose grism G600B to cover the region of Ly$\alpha$ at z=3.036
and grism G600R to look for other emission lines at longer wavelengths.
Even though we did not expect to see any lines in the red spectra, it
is imperative that they are deep enough that we can rule out the
alternative interpretation that our objects are \ion{O}{ii} emitters
rather than Ly$\alpha$ emitters. G600B has a wavelength coverage from
3600\AA \ to 6000\AA \ (depending somewhat on the position on the CCD) 
and a spectral resolution of 815 whereas G600R covers the range 
5200\AA--7400\AA \ at a spectral resolution of 1230.

We constructed three independent masks. In addidition to covering
all eight candidate LEGOs, several of them in more than one mask,
this also allowed us to obtain spectra of objects close to the quasar
line of sight. For the mask construction we used the
{\it FORS Instrumental Mask Simulator} (FIMS). We hereafter refer to the
three G600B masks as maskB1, maskB2 and maskB3, and to the three G600R 
masks as maskR1, maskR2 and maskR3. 

\begin{table}
\begin{center}
\caption{The log of VLT observations.}
\begin{tabular}{@{}lllcccc}
\hline
date        & setup & seeing & Exposure time \\
            &       & arcsec &        (sec) \\
\hline
\hline
Imaging : &          &          &      \\
2000 Jan 12, 17 & Bessel B & 0.68-1.02 & 5200 \\
2000 Mar 5 & Bessel I & 0.55-0.75 & 3750 \\
\hline
Spectroscopy : &          &          &      \\
2000 Mar 5 & MOS, maskB1 & 0.78--0.96 & 7200 \\
2000 Mar 4 & MOS, maskB2 & 0.78--1.18 & 7200 \\
2000 Mar 4 & MOS, maskB3 & 0.61--0.66 & 7200 \\
2000 Mar 4 & MOS, maskR1 & 0.77--0.93 & 5400 \\
2000 Mar 5 & MOS, maskR2 & 0.57,0.78  & 3600 \\
2000 Mar 5 & MOS, maskR3 & 0.59--0.95 & 5400 \\
\hline
\label{obs-journal}
\end{tabular}
\end{center}
\end{table}

\subsection{Spectroscopy and I-band imaging}
The spectroscopic observations were carried out during the two nights
of March 4--5 2000 under photometric and good seeing conditions. We 
obtained total integration times of 4$\times$1800sec for each of
maskB1--maskB3 and 3$\times$1800sec, 2$\times$3600sec and
3$\times$1800sec for maskR1, maskR2 and maskR3 respectively. As the
seeing fwhm for all frames was significantly smaller than the
slit-width the spectral resolution is determined by the seeing and/or
the size of the object.
We also obtained 15$\times$250sec images in the Bessel 
I--band filter. During the G600B observations the CCD was binned 
2$\times$2 in order to reduce the influence of read-out noise. For all 
spectroscopic observations the CCD was read out in single port, high 
gain mode. For the imaging observations the CCD was read out using the 
four port, high gain mode.  
The journal of observations appears as Table~\ref{obs-journal}.

\begin{table*}
\begin{center}
\caption{Properties of the confirmed LEGOs S7--S13. Magnitudes
are measured in a 3.5 arcsec diameter aperture. Two Ly$\alpha$ fluxes
are given, one as measured through the slit (flux(slit)) and one
corrected to the 3.5 arcsec aperture (flux(aper)
see text for details). The errors on
the linefluxes are propagated errors from photon statistics relevant
for determining the significanse of the detection, they do not include
calibration errors which are estimated to be 30\% because of the large
aperture correction.}
\begin{tabular}{@{}lccccccccr}
\hline
Source  & wavelength & redshift & fwhm & velocity width & B mag & I mag
& flux(slit)$\times 10^{17}$ & flux(aper)$\times 10^{17}$ & W$_{\rm obs}$ \\
        &  (\AA)     &          &  (\AA)        &   km s$^{-1}$   &    &
	& erg s$^{-1}$ cm$^{-2}$ & erg s$^{-1}$ cm$^{-2}$ & (\AA) \\
\hline
S7  & 4911.6 & 3.0402 & 7.4 & $<$350 & 25.61$\pm 0.11^a$ & 24.28$\pm 0.09^a$ &
1.04$\pm$0.15 & 1.58$\pm$0.23 & $>61~$\, \\ 
S8  & 4911.1 & 3.0398 & 6.5 & $<$240 & $\sim$27.3$^b$ & $\sim$26.8$^b$ &
1.70$\pm$0.18 & 2.58$\pm$0.27 & $456^b$ \\
S9  & 4905.2 & 3.0350 & 7.2 & $<$340 & 25.36$\pm$0.07 & 24.61$\pm$0.13 &
3.51$\pm$0.18 & 5.34$\pm$0.27 & $>164~$\, \\
S10 & 4905.6 & 3.0353 & 7.2 & $<$240 & $\sim$27.3$^b$ & $\sim$26.8$^b$ &
2.09$\pm$0.26 & 3.17$\pm$0.40 & $456^b$ \\
S11 & 4900.6 & 3.0312 & 7.2 & $<$440 & $\sim$27.3$^b$ & $\sim$26.8$^b$ &
1.48$\pm$0.26 & 2.25$\pm$0.40 & $456^b$ \\
S12 & 4903.2 & 3.0333 & 8.2 & $<$520 & $\sim$27.3$^b$ & $\sim$26.8$^b$ & 
1.54$\pm$0.31 & 2.34$\pm$0.48 & $456^b$ \\
S13 & 4890.4 & 3.0228 & 7.3 & $<$450 & $\sim$27.3$^b$ & $\sim$26.8$^b$ & 
1.39$\pm$0.19 & 2.11$\pm$0.29 & $456^b$ \\
\hline
\label{redtab}
\end{tabular}
\end{center}
\begin{flushleft}
\begin{footnotesize}
\vskip -0.4cm
\noindent
$^a$ Including the nearby red object to the West of S7 proper

\noindent
$^b$ Measured in the median image of S8, S10--S13
\end{footnotesize}
\end{flushleft}
\end{table*}

\subsection{Data reduction}
   The pre--imaging B--band images and the I--band images were BIAS 
subtracted and flat--fielded 
using standard techniques and the individual reduced images were 
combined using the $\sigma$--clipping algorithm described in M\o ller 
and Warren (1993).  The combined images reach 5(2)$\sigma$ detection 
limits in 1 arcsec$^2$ circular apertures of 25.9(26.9) in the I--band 
and 26.7(27.7) in the B--band (both in the AB-system). 

The spectroscopic frames were first BIAS subtracted. The 
flat-fielding was thereafter done in the following way. First we median 
filtered
the flat-fields along the dispersion direction (x-axis) using a 
30$\times$1 boxcar filter for the G600B flat-frames and a 
60$\times$1 boxcar filter for the G600R flat-frames. Then we 
normalised the flat-frames by dividing the unfiltered 
flat-frames with the filtered flat-frames. Finally, we
divided the science frames with these normalised flat-frames. 
   The individual BIAS subtracted and flat-fielded science
spectra were subsequently sky-subtracted in the following
way. First we removed cosmic ray hits from the sky-region
of the 2-dimensional spectrum by $\sigma$-clipping along
each spatial direction column. We determined 
a 2-dimensional sky-frame using the {\it background} task 
in the {\it kpnoslit} package in IRAF. This sky-frame was 
then subtracted from the original BIAS subtracted and 
flat-fielded 2-dimensional science spectrum. The individual 
reduced and sky-subtracted science spectra were then combined 
using $\sigma$-clipping for rejection of cosmic ray hits. 

1-dimensional spectra were extracted using the {\it apall}
task. For objects without or with very faint continuum 
we used bright objects from neighbour slitlets to determine
the trace of the spectra.

The 1-dimensional spectra were wavelength calibrated 
using the {\it dispcor} task. The rms of the deviations
from the fits to 3. order chebychev polynomia were 0.3--0.5\AA 
\ for the G600B spectra and 0.05--0.1\AA \ for the G600R
spectra. The high rms values for the wavelength calibration
of the G600B spectra is due to the binning of CCD for the
G600B spectra which imply fewer pixels per resolution element.

\section{Results}

\subsection{Spectroscopy}

\subsubsection{Emission line candidates and line fluxes}
\label{sec:Emis}
Our candidate list consisted of six ``certain'' candidates
(Ly$\alpha$ S/N$>$5) named S7--S12 and two ``possible'' candidates
(Ly$\alpha$ S/N$\approx$4) named S13 and S14. The candidates were
selected on the basis of excess flux in the narrow--band as compared
to their B and I band magnitudes (see paper I). A Ly$\alpha$ emission
line was detected spectroscopically for all objects except S14. The
non--confirmation of S14 is consistent with the new and deeper B--band
image also obtained during this run (Section \ref{sec:Img}), as the
detection of B--band continuum for S14 lowered the significanse of the
narrow--band excess below 3$\sigma$.

\begin{figure*}
\begin{center}
\epsfig{file=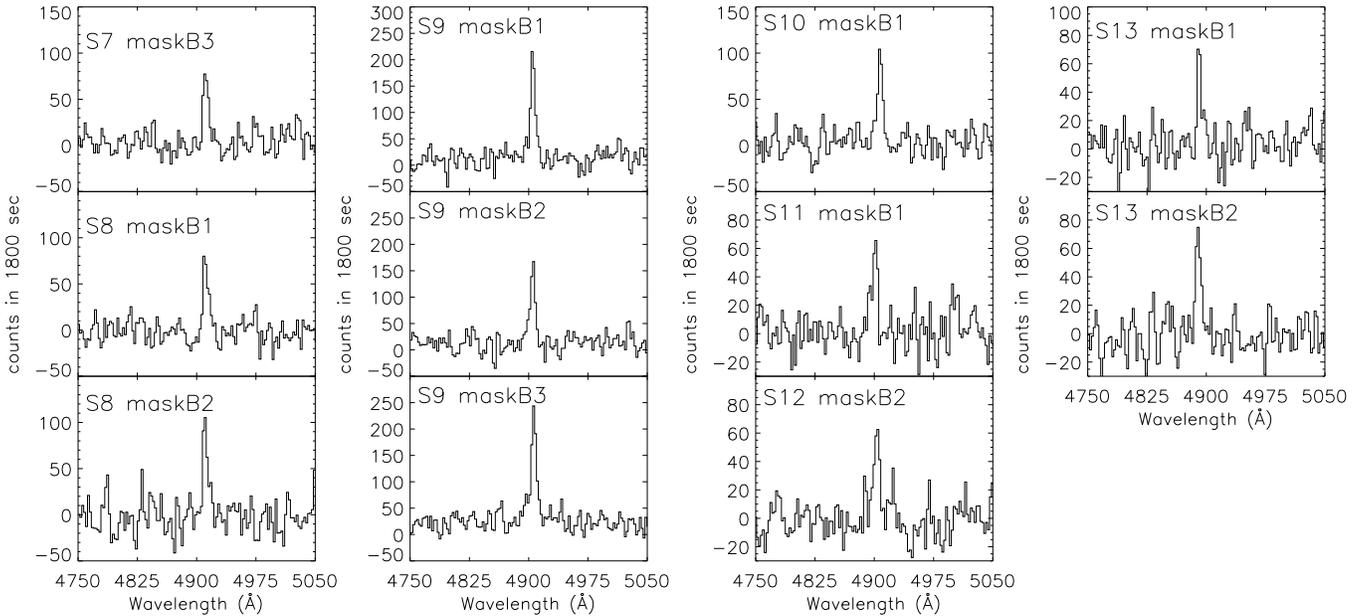,width=18cm}
\caption{The spectral regions around the Ly$\alpha$ emission lines for the
LEGOs S7--S13. For S8, S9, and S13 we show the spectra from several
masks. As seen, the presence of emission lines is confirmed for
all 7 candidates. For S14 (not shown) we did not detect an
emission-line within the range of the narrow filter.
}
\label{Sspec}
\end{center}
\end{figure*}

In Fig.~\ref{Sspec} we show extractions of the blue grism spectra
of all the candidates for which we confirm the presence of an emission 
line within the transmission curve of the narrow filter. 
Only for S7 and S9 did we detect a faint continuum in the spectra. 

With variable seeing and 1.2 arcsec slits it was not possible to obtain
a spectrophotometric calibration. Nevertheless we obtained a rough
calibration as follows. For each mask we had placed several slits on
stars for which we had accurate B and I magnitudes. For the G600B
masks we selected a star with a flat spectrum in B, and used that as
a local standard for the flux calibration of emission lines. The
calibration was done for each mask individually, and the results were
combined afterwards for those objects that were observed more than once.
The scatter for the objects observed more than once was about two--three
times larger than the observational errors, confirming that this
calibration is dominated by slit--losses as we expected. The resulting
line fluxes are given in Table~\ref{redtab} (flux(slit)).

The flux calibration described above is only valid if
all the emission line objects are point sources. For
extended objects the slit--losses will be larger than for the standard
star, and an additional aperture correction must be applied. Comparing
to the imaging line fluxes found with a circular aperture of diameter
3.5 arcsec (paper I), we find a mean aperture correction of 0.46 mag.
Line fluxes including this mean aperture correction are also given
in Table~\ref{redtab} (flux(aper)).

\subsubsection{Identifications and redshifts}
Before we can conclude that the emission lines are indeed due to
Ly$\alpha$ at z$\approx$3.04, we must first exclude any possible low
redshift ``contamination''. In the deep narrow band imaging discussed
by Kudritzki et al. (2000) 9 of 10 candidates were confirmed as
Ly$\alpha$ emitters at z=3.1, while the last proved to be
a galaxy at z=0.35 where the \ion{O}{ii} 3727{\AA} line fell in the
same narrow--band filter. There are other, albeit less likely,
possible identifications (\ion{Mg}{ii} and \ion{Ne}{iii}). Those can 
all be tested by searching for additional emission lines (\ion{O}{ii}, 
H$\beta$ and especially \ion{O}{iii}). 
For most of the possible interpretations the
\ion{O}{iii} would be redshifted out of the G600B spectra. For this
reason we also obtained spectra of all objects with the G600R
grism.

\begin{figure}[t]
\begin{center}
\epsfig{file=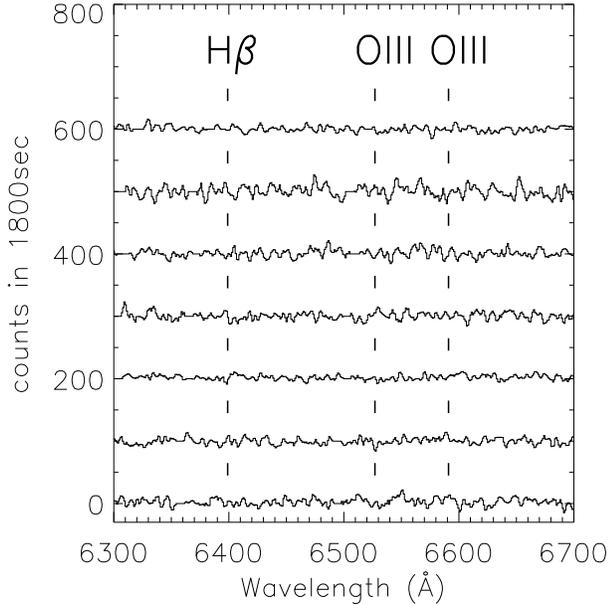,width=9cm}
\caption{Shown here are parts of the G600R spectra were H$\beta$ and 
\ion{O}{III} lines would have been if the emission line detected with
the narrow--band imaging (paper I) were due to \ion{O}{II} at z=0.31.
S7 is at the bottom and S13 at the top. The spectra have been smoothed
with a seven pixel (7.4\AA) box car filter. For S8, S9 and S13 the
spectra are the average of 2, 3, and 2 spectra respectively.
}
\label{Sred}
\end{center}
\end{figure}

For none of the confirmed emission--line objects S7--S13 did we
detect any other emission line, neither in the G600B nor in the
G600R spectra. In Fig.~\ref{Sred} we
show the regions of the G600R spectra where the H$\beta$
and \ion{O}{iii} lines would have fallen if the emission lines
seen in Fig.~\ref{Sspec} had been \ion{O}{ii} at z=0.313 (S7
at the bottom and S13 at the top). 
The spectra have been smoothed by a 7 pixel (7.4\AA) boxcar filter 
and regions with large errors due to strong sky-lines were set to
zero. As seen, H$\beta$ and \ion{O}{iii} lines are not 
present in any of the spectra.
The question of possible alternatives to the Ly$\alpha$
identification is an important one, and we shall return to a complete
discussion in Section \ref{sec:ident}. For now we shall assume that
the lines are indeed due to Ly$\alpha$.

The wavelengths and widths of the emission lines were determined by
fitting them with Gaussian profiles. The uncertainty in the
determination of the wavelength centroid is about 0.3{\AA}. The
results are given in Table~\ref{redtab} where we also list the
resulting redshifts under the assumption that the lines are due to
Ly$\alpha$.

In order to constrain the intrinsic width of the lines we must know the
spectroscopic resolution. An upper limit to the fwhm of the resolution
profile along the dispersion direction can be obtained from the 
width of the slitlets and the dispersion. With a slit width of 1.2
arcsec (3 pixels) and dispersion of 2.4\AA \ per pixel the resolution
would have been about 7.2\AA \ if the seeing had been worse than 1.2
arcsec. However, as the seeing was in all cases significantly better
than 1.2 arcsec the spectroscopic resolution is smaller than 7.2\AA \
fwhm. An upper limit to the spectroscopic resolution can be derived from
the spatial profile of the spectra of point sources at wavelengths near
4900\AA. Converted to \AA \ the spatial widths are 5.9\AA, 5.7\AA \ and 
4.8\AA \ for maskB1, maskB2 and maskB3 respectively. As the instrumental
resolution for FORS is slightly lower along the dispersion direction
than along the spatial direction (T. Szeifert, private communication)
these values can be used as lower limits on the spectroscopic
resolution. From this lower limit we can obtain upper limits to the
intrinsic widths of the lines by deriving the intrinsic widths
that convolved with a Gaussian with a width corresponding 
to the lower limit on the spectroscopic resolution reproduces the 
observed line widths. These upper limits are also given in
Table~\ref{redtab}. The line widths are smaller than 6\AA \ or 370 km
s$^{-1}$ for Ly$\alpha$ at z=3.04.

\begin{figure*}[ht]
\begin{center}
\epsfig{file=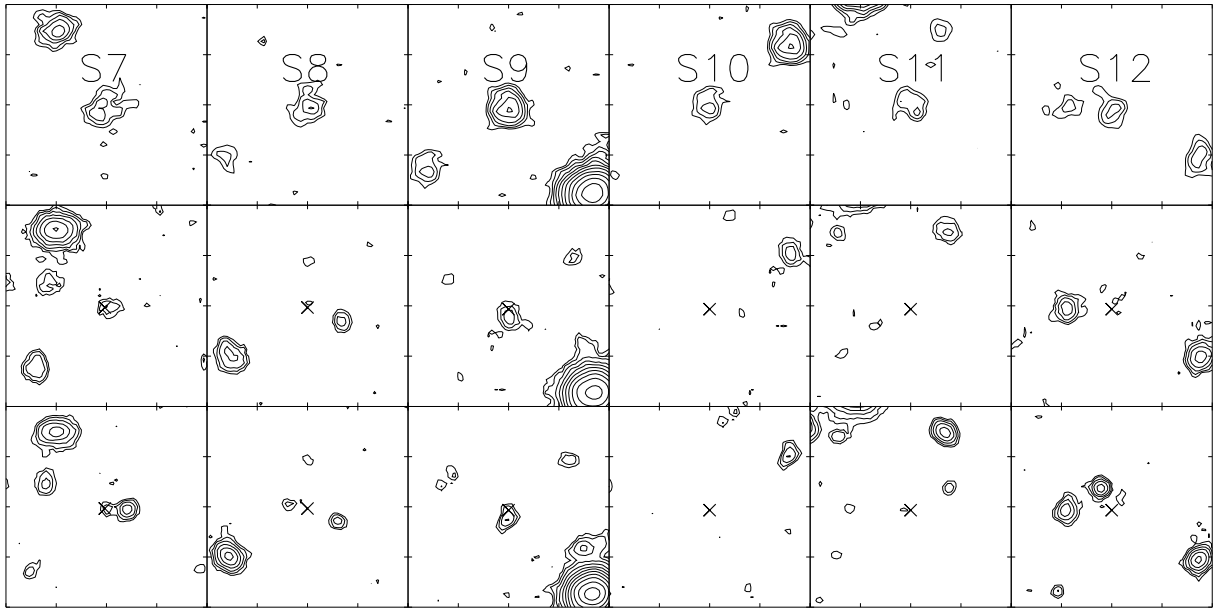,height=8cm}
\caption{Contour plots of the combined NTT narrow--band (top), VLT 
B--band (middle) and VLT I--band (bottom) images for each of the six 
candidate emission line galaxies S7--S12 detected with S/N$>5$ in the 
NTT narrow--band image (paper I). The size of the individual fields 
is 12$\times$12 arcsec$^{2}$ and the fields are orientated with east
to the left and north up. The contour levels are logarithmic. 
}
\label{Scont}
\end{center}
\end{figure*}

\subsection{Imaging}
\label{sec:Img}
\begin{figure}[ht]
\begin{center}
 \epsfig{file=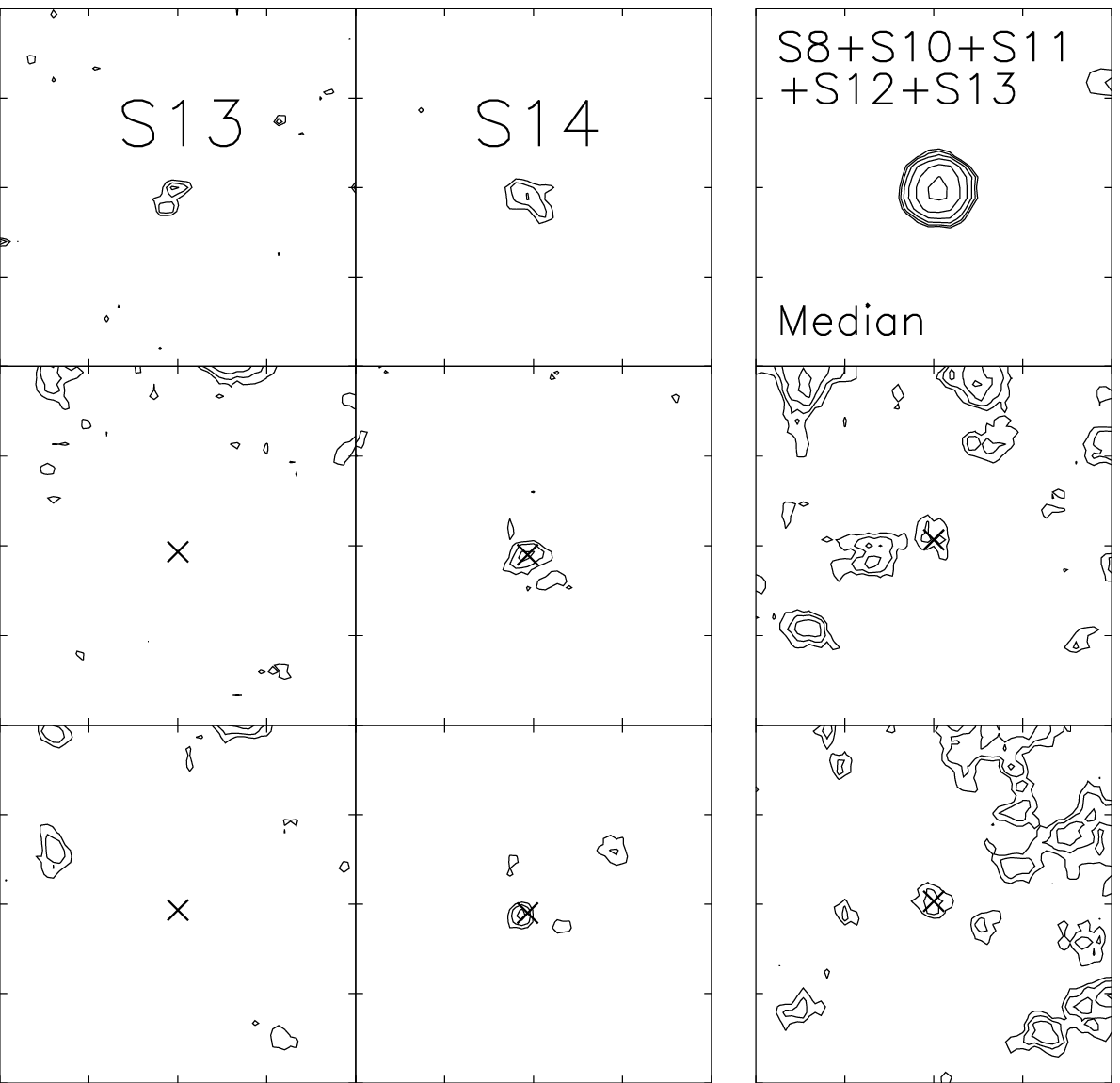,height=8cm}
 \caption{{\it Two left panels:} Contour plots of the combined NTT 
narrow--band (top), VLT 
B--band (middle) and VLT I--band (bottom) images for the two 
candidate emission line galaxies S13 and S14 detected with $S/N<5$ 
in the NTT narrow--band image. The size of the individual fields 
is 12$\times$12 arcsec$^{2}$ (as in Fig.~\ref{Scont}) {\it Right
panel:} Same as Fig.~\ref{Scont} and the two left panels, but for
the median image of S8, S10--S13 and with different contour levels. 
Broad band emission is detected in the median images at the exact
position of the stacked narrow--band object. 
}
\label{SN4}
\end{center}
\end{figure}

Fig.~\ref{Scont} and the two left panels in Fig.~\ref{SN4} show
regions of size 12$\times$12 arcsec$^{2}$ centred on each of the
objects S7--S12 and S13--S14 from the combined NTT narrow--band image
(top row), VLT B--band image (middle row) and VLT 
I--band image (bottom row). Only two of the objects, S7 and S9, are
detected above our 2$\sigma$ detection limits of I(AB)=26.9 and
B(AB)=27.7 (detection limits for 1 arcsec$^2$ circular apertures).
Those two galaxies were already detected in the deep NTT
images presented in paper I. For S9 both the B--band and I--band
emission is centred 0.51$\pm$0.05 arcsec south of the narrow-band
position. In the broad band images of S7 we see two nearby objects.
The center of the Ly$\alpha$ emission is found 0.21$\pm$0.11 arcsec
east of the easternmost of the two, which we identify as the likely
host of the Ly$\alpha$ emission. We cannot be certain that the
nearby western object is unrelated, hence the aperture magnitudes
for S7 given in Table~\ref{redtab} includes both of the objects.
The not--confirmed ($<5\sigma$) candidate S14 is clearly detected in
both the B and I bands. The B magnitude is bright enough to confirm that
the apparent narrow--band excess in the old data set was not
significant.

For the five remaining objects we detect no broad band flux above
2$\sigma$. In order to constrain the limit on the broad band emission
further we registered the sub--images of the objects S8, S10--S13 to
the centroids of their Ly$\alpha$ emission and coadded them.
In the right panel of Fig.~\ref{SN4} we show the median of the
coadded images. We now detect a faint object in both the I--band
and the B--band. To be able to compare the fluxes we apply aperture
corrections determined from a bright point source and arrive at
I(AB) = 26.8$\pm$0.3 and B(AB) = 27.3$\pm$0.2 for the
3.5 arcsec diameter circular aperture.

If this flux was due mainly to one or two of the objects, they
would have been visible on the individual images. Hence, we conclude
that the flux must be fairly evenly distributed on most or all of the
five objects, and that they each must have roughly the magnitudes measured
in the combined images. It is interesting to note that unlike S7 and S9,
the continuum emission in the coadded frames is centered on the same
position as the Ly$\alpha$ emission. This suggests that the relatively
bright continuum object identified as S9 and the two--component
object identified as S7 are, at least partly, due to a chance alignment
of unrelated objects. Hence, the broad--band magnitudes given for S7 and 
S9 should probably be regarded as upper limits on their brightness.

M{\o}ller and Warren (1998) reported that on HST images of three
Ly$\alpha$ emitting objects related to the DLA at z=2.81 in front of
PKS0528--250, there was evidence that the Ly$\alpha$ emission was
significantly more extended than the continuum sources. We measured
the FWHM of the Ly$\alpha$ emission of S9 and of the stacked Ly$\alpha$
object, and found an intrinsic size (after correcting for the seeing)
of 0.98 and 0.65 arcsec (FWHM). This supports the result reported
by M{\o}ller and Warren (1998).

\section{Discussion}

\begin{figure*}[ht]
\begin{center}
 \epsfig{file=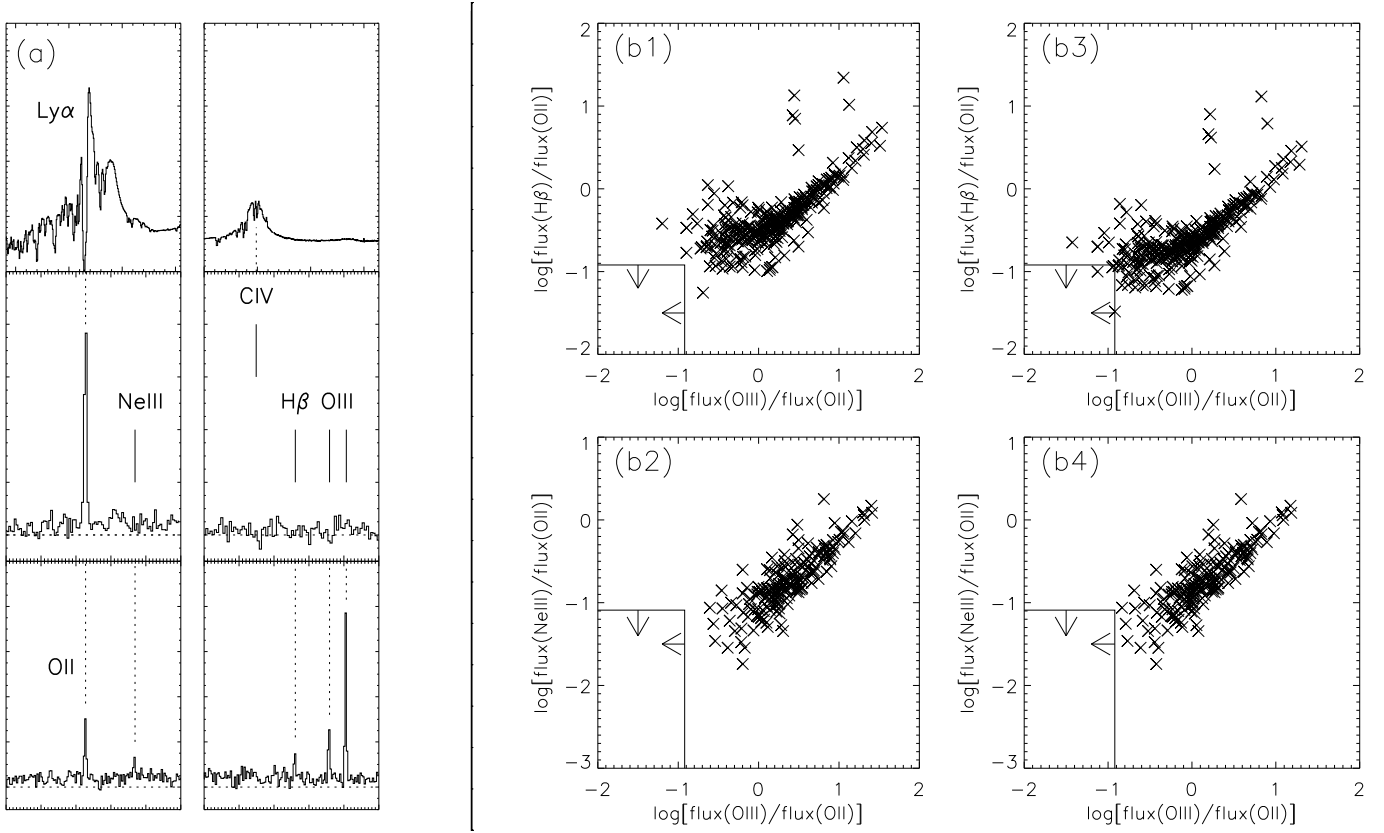,height=11cm}
\caption{
{\bf a:} Middle panel shows the stacked G600B (left) and G600R (right)
spectrum of S7--S13. Lower panel shows, for comparison, the spectrum of 
a z=0.224 emission line galaxy redshifted to
z=0.313 so that the \ion{O}{ii} line falls at the wavelength of the
observed emission line of S7--S13. The stacked spectrum of
S7--S13 has none of the lines expected if S7--S13 had been foreground
emission line galaxies. Upper panel shows the spectrum of Q1205-30
to indicate the position of the \ion{C}{IV} emission expected for AGNs. 
A \ion{C}{IV} emission line is not detected. 
{\bf b:} The box to the lower left of each of the figures
{\it (b1,2,3,4)} marks the 2$\sigma$ limits for the flux ratios  
$\log(f_{\rm OIII}  /f_{\rm OII})$, $\log(f_{\rm H\beta}/f_{\rm OII})$, 
and $\log(f_{\rm NeIII} /f_{\rm OII})$ under the assumption that S7--S13
are z=0.313 galaxies with \ion{O}{ii} in the narrow filter. The
observed flux ratios from the sample of Terlevich et al. (1991)
are plotted ($\times$'s) under two assumptions used for the conversion
of equivalent widths to fluxes (``mean'' for b1 and b2, and
``worst case'' for b3 and b4 respectively, see text for details).
Under both assumptions the 2$\sigma$ limits firmly exclude the
z=0.313 hypothetis.
}
\label{Terl}
\end{center}
\end{figure*}

\subsection{Line identification}
\label{sec:ident}

We now return in more detail to the identification of the detected 
emission lines.  For a discussion of ways to discriminate between Ly$\alpha$ 
and lower redshift objects for single emission lines around 8000--9000\AA, 
corresponding to very high redshifts (z$\gtrsim$5) for Ly$\alpha$, see
Stern et al. (2000). Here we discuss possible 
contaminants for z=3 Ly$\alpha$ emitters.

\ion{Mg}{ii} and \ion{Ne}{iii} can both be excluded easily since
in this case we would have detected the stronger \ion{O}{ii} line.
In the following we will discuss how to reject the possibility
that the observed line is due to \ion{O}{ii}.
There are two independent ways to
check the likelyhood of the identification as Ly$\alpha$ rather than
\ion{O}{ii}. One is by
its equivalent width, the other by upper limits on line flux ratios.
We shall here apply both tests, and as reference point for the low
redshift identification we have chosen the large samle of emission
line galaxies from the survey of
Terlevich et al. (1991).

\subsubsection{Equivalent widths}
The expected rest equivalent width of unextinguished star formation
induced Ly$\alpha$ can be as high as 200--300\AA\ (Charlot and Fall
1993; Valls--Gabaud 1993), which at a redshift of 3 would result in
an observed equivalent width of order 1000{\AA}. Even
with a fair amount of absorption by dust, the observed equivalent width
would remain large.

The continuum of low redshift \ion{O}{ii} galaxies is, in contrast to
that of the Ly$\alpha$ selected galaxies at high redshift,
for the most part easily detected. In the sample of Terlevich et al.
(1991) we find a median \ion{O}{ii} rest equivalent width of 59\AA , and
the 95\% quantile is 163\AA .
In Table~\ref{redtab} we list the observed equivalent widths
(W$_{\rm obs}$) of our lines. As discussed above, the broad--band
aperture magnitudes of the two objects S7 and S9 probably include
flux unrelated to the emission line objects, and hence the calculated
equivalent widths are lower limits. For the remaining objects it was
necessary to coadd the images before we were able to measure their
broad--band flux. For consistency we list the equivalent width measured
on the stacked images.

For an \ion{O}{ii} galaxy at z=0.31 we would expect a median
equivalent width of 77\AA\ and none of seven objects should be above the
95\% quantile of 214\AA . This is incompatible with the observed
distribution.

\subsubsection{Line intensity ratios}

In order to compare the intensity ratios, we first had to determine
the upper limit for the non--detection of lines in the predicted
region for \ion{O}{iii} and H$\beta$. For the flux calibration
of the G600R spectra we used the exact same method as described in
Section \ref{sec:Emis} for G600B. To maximixe the S/N for the
non-detection of H$\beta$ and \ion{O}{iii} emission lines, we
stacked all the spectra. In Fig 5(a) we show the stacked spectrum
(middle panel), and we have marked the expected positions of various
emission lines. We find 2$\sigma$ upper limits to the log of flux
ratios as follows:
log($f_{\rm OIII}  /f_{\rm OII}) < -0.92$,
log($f_{\rm H\beta}/f_{\rm OII}) < -0.92$, and
log($f_{\rm NeIII} /f_{\rm OII}) < -1.09$.
Above the stacked spectrum we plot the spectrum of the quasar
in order to show the positions of the Ly$\alpha$ and \ion{C}{iv}
emission lines. No \ion{C}{iv} line is detected in the stacked
spectrum. The flux ratio
(log($f_{\rm CIV} /f_{\rm Ly\alpha}) < -0.92$)
excludes a significant AGN contribution to the Ly$\alpha$ emission.
Below the stacked spectrum we show for comparison the spectrum of an
\ion{O}{ii} galaxy observed by chance during the same run
in one of the slits we placed on random objects in order to fill the
masks. This galaxy was observed at a redshift of z(\ion{O}{ii})=0.224,
but we have shifted it to z=0.313 for easy comparison.

In Terlevich et al. (1991) we only found data for the equivalent widths
of their emission line sample. In order to convert those to intensity 
ratios of two lines, we need first to determine the ratio of the local 
continuum flux under those two lines. In all the published spectra of 
the Terlevich sample we therefore measured the ratio of the continuum 
flux at 3727--3870\AA \ (OII and NeIII) and at 5007\AA \ (OIII). We then 
determined the mean of this ratio and the maximum of the 
ratio (which is the worst case), and used both of them to
convert the equivalent width ratios to flux ratios. The results are plotted
in Fig. 5(b). In Fig. 5(b1,b2) we plot line flux ratios calculated
from the mean continuum slope. The box and arrows in the
lower left corner marks the 2$\sigma$ upper limits of our
non--detections. In Fig. 5(b3,b4) we repeat the same plots, but here
we have used the ``worst case'' continuum for the conversion to flux
ratios. Even in this case our 2$\sigma$ upper limits have no overlap
with the \ion{O}{ii} galaxy distribution.

In summary of this section we have applied two independent methods
(equivalent widths and line flux ratios) to test how well our sample
of emission line objects would fit if interpreted as low redshift
\ion{O}{ii} galaxies. Both tests reject the interpretation, and
we conclude that all seven objects
presented here are Ly$\alpha$ emitters at z$\approx$3.04.

\subsection{Star--formation rates}

In paper I we calculated star--formation rates (SFR) based on the
Ly$\alpha$ fluxes. Here, for comparison, we calculate the SFRs from
the continuum fluxes. As detailed in Section \ref{sec:Img} we have
reasons to believe that the continuum of S7 and S9 may be boosted by
neighbour objects, therefore we perform the calculation for the
remaining five objects only.

The restframe UV continuum in the range 1500\AA--2800\AA \
can be used as a SFR estimator if one assumes that the star-formation is
continuous over a time scale of more than 10$^8$ years. Kennicutt (1998)
provides the relation
\[
SFR(M_{\sun} yr^{-1}) = 1.4 \times 10^{-28} \times L_{\nu},
\]
where $L_{\nu}$ is the luminosity in the 1500\AA--2800\AA \ range
measured in erg s$^{-1}$ Hz$^{-1}$. The observed I--band corresponds to
the rest-frame UV continuum around 2000\AA, which falls well within
this range. We can hence use the I--band flux in the median image
of S8, S10--S13 as a measure of their SFRs. 
To derive $L_{\nu}$ we used the definition of the AB magnitude to derive
the observed flux ($F_{\nu} = 10^{-0.4\times(I(AB)+48.6)}$) and finally
the luminosity distance in our assumed cosmology
(d$_{lum}$=8.58$\times$10$^{28}$cm) to derive $L_{\nu}$ :
\begin{eqnarray}
L_{\nu} &=& F_{\nu} \times 4 \pi d_{lum}^2/(1+z) {}
\nonumber\\
&=& {} 1.5 \times 10^{28}\:erg s^{-1} Hz^{-1} {},
\nonumber
\end{eqnarray}
where the factor $(1+z)^{-1}$ corrects for the fact that $L_{\nu}$ is
a specific luminosity (not a bolometric luminosity). Using the relation
of Kennicutt (1998) we find a SFR of 2.1 M$_{\sun}$ 
yr$^{-1}$. This is 1--2 orders of magnitude smaller 
than the SFRs derived for LBGs (Pettini et al. 1998) and 2--3 orders of
magnitudes smaller than for the sub--mm selected sources (Ivison et 
al. 2000).

From the Ly$\alpha$ fluxes we found SFRs for the same objects in the
range 0.3--0.5 $h^{-2}$ M$_{\sun}$ yr$^{-1}$ for a $\Lambda = 0$
cosmology. Converting to the cosmology used in this paper this
corresponds to 1.6--2.6 M$_{\sun}$ yr$^{-1}$, which is
identical to what we now find from the I--band flux. This result is
inconsistent with the presence of large amounts of dust in those
objects, which is not too surprising as one would expect that a
targeted search for Ly$\alpha$ emitters would preferentially find
objects with very little or no dust. This {\it does} however imply
that a large number of small star--forming objects at high redshift have
essentially no dust in them.

\subsection{Space density}

The density of Ly$\alpha$ emitters derived from the seven confirmed
objects is 16$\pm$4 per arcmin$^2$ per unit redshift (seven objects
within the 27.6 arcmin$^2$ field of view of the EMMI instrument of NTT
and within the $\Delta z$=0.016 range of the narrow filter). 
In Table~\ref{surveys} we compare this to the results from
other recent searches for LEGOs at z$\approx$3. It is seen from
Table~\ref{surveys} that our survey found a larger space density than
any other survey, but also that we reach the faintest detection
limit of them all. The known LBG overdensity studied by Steidel et al.
(2000) has a similar space density, but to a three times brighter 
flux limit. Down to their flux limit we would only have detected two
of our seven sources. It is hence not clear from this if the volume
around Q1205-30 is overdense in LEGOs, or if we simply see the
effect of observing to a lower limiting flux. To be able to address
this question, we need to compare our results to the extrapolation
of the LBG LF.

\begin{table*}[t]
\begin{center}
\caption{The properties of other recent searches for LEGOs at
z$\approx$3.}
\begin{tabular}{@{}lllccccccc}
\hline
Survey  & z,$\Delta$z & Area & 5$\sigma$ limit & N & $\frac{dN}{d{\Omega}dz}$
& field & Confirmed & Ref \\
        &             & $\square$' & $\times$10$^{-17}$ erg s$^{-1}$ cm$^{-2}$

& & \#/$\square$' & \\
\hline
PKS0528-250 & 2.81, 0.019 & 27 & 3.7 & 3 & 5.9$\pm$3.4 & QSO, DLA & all
& (1,2) \\
HDF N   & 3.43, 0.063 & 29   & 3.0 & 5 & 2.7$\pm$1.2 & blank & all, 2
AGN & (3,4) \\
SSA 22  & 3.43, 0.063 & 30   & 1.5 & 7 & 3.7$\pm$1.4 & blank & all & (3,4)
\\
BR0019-152 & 3.43, 0.063 & 16 &  -  & 7 & 6.9$\pm$2.6 & blank & 3/7 &
(3,4) \\
Virgo & 3.15, 0.043 & 50 & 2.0 & 9 & 4.2$\pm$1.4 & blank & all & (5) \\
SSA 22a & 3.09, 0.066 & 78 & 3.0$^a$ & 72 & 14$\pm$1.6 & LBG spike & 12/72
& (6) \\
Q1205-30 & 3.04, 0.016 & 28 & 1.1 & 7 & 16$\pm$4 & QSO & all & (7) \\ 
\hline
\label{surveys}
\end{tabular}
\end{center}
\begin{footnotesize}
\vskip -0.6cm
\noindent
$^a$ The true 5$\sigma$ level for this study is fainter, but only
objects brighter than this limit were included in the catalog (Pettini,
private communication)
\end{footnotesize}
\vskip 0.1cm
References : (1) M\o ller and Warren (1993); Warren and M\o ller (1996);
(3) Hu et al. (1998); (4) Cowie and Hu (1998);
(5) Kudritzki et al. (2000); (6) Steidel et al. (2000);
(7) Fynbo et al. (2000) and this paper.

\end{table*}

Adelberger and Steidel (2000)
present a LF for LBG selected galaxies which is based
on the HDF--North for the faint (25$<$R$<$27) and ground based 
LBG surveys for the bright (R$<$25.5) end. Integrating this
LF down to R=27 leads to a predicted density of
objects of 0.017 Mpc$^{-3}$h$^{3}$. The comoving volume probed by 
our survey is 5.8$\times$10$^2$ Mpc$^{3}$h$^{-3}$. Based on the 
LF of LBGs we therefore expect ten galaxies in the
volume and we find seven. However, only $\sim$20\% of the Lyman-Break
galaxies show Ly$\alpha$ in emission (Steidel et al. 2000). This may
be an underestimate if Ly$\alpha$ and continuum emission in general
have different spatial distributions due to different slit--losses.

With the Lyman--Break technique the only way to probe the R(AB)$>$25.5
part of the LF is to use the Hubble Deep Fields. This
makes it impossible at present to study the faint end LF of any
significant volume with this technique, and it is therefore very
uncertain. Indeed, significant differences in the numbers of high redshift
objects in the HDF North and South fields based on photometric redshifts
have been reported (Fontana et al. 2000).
The faint end of the LF is, however,
important to study because a significant fraction of the
star--formation, and therefore also the background ionising photons, may
originate there. Based on the LF of Adelberger and
Steidel (2000) there are roughly equal amounts of total luminosity from
galaxies with R$<$25.5, 25.5$<$R$<$27 and 27$<$R$<$30. Hence, even if
this LF is correct, only about one third of the total
star formation at z=3 is traced by the LBGs studied in the ground based
samples. In case the faint end LF needs to be revised upwards it will
be even less.

Assuming for now that the 20\% Ly$\alpha$ fraction can be extrapolated
to the faint end of the LF, we find that the volume we have surveyed
has a comoving density of faint LBGs which is 3.5 times that predicted
by the HDF--North LF. This could indicate that at least some radio quiet
QSOs reside in overdense environments. Another likely interpretation is, 
however, that the faint end of the LF has a larger fraction of Ly$\alpha$
emitters, in our case it must be of order 70\% to fit the HDF--North
LBG LF. A physical explanation could be that the objects in the
faint end of the LBG LF are less dusty that the bright LBGs.

\subsection{Redshift distribution}

The Ly$\alpha$ emitters were selected in a deep narrow--band search. It
now remains to be tested if the objects span the entire width of the
narrow--band filter, or if they cluster within a wavelength range
smaller than the filter width. The width of the filter response was
23{\AA}, and the full width spanned by the seven objects is 21.2{\AA}.
Monte Carlo simulations, where we randomly distributed seven objects
weighted by the filter transmission curve and measured the resulting
mean and std.dev., show that the seven redshifts of S7--S13 are
consistent (to within 1$\sigma$) with being drawn from a random
distribution. Therefore, the structure we have found is most likely
larger than the redshift span we have covered with our filter.

\section{Conclusions}

We have reported on spectroscopic observations of eight candidate
Ly$\alpha$ emitting objects at z$\approx$3.04. All six ``certain''
($>5\sigma$) candidates were confirmed, and of the two ``possible''
($<5\sigma$) candidates one was confirmed. To assess the most likely
identification of the lines we performed two independent detailed
tests based on a large samle of low redshift \ion{O}{ii} galaxies.
Both tests indicate that the only likely identification is Ly$\alpha$.
This conclusion is strengthened by the fact that the original
narrow--band imaging was centered on a quasar at the same redshift.
The seven Ly$\alpha$ objects are hence likely associated with the same
structure as the quasar.

The narrow--band images of the objects, as well as the large
slit--losses of Ly$\alpha$ emission
we need to correct for when calibrating the line fluxes,
indicate that the Ly$\alpha$ emission originates in extended objects.
This is similar to the results reported by M{\o}ller and Warren (1998)
on Ly$\alpha$ emission related to a DLA at z=2.81, where it was found
that the Ly$\alpha$ emission was significantly more extended than the
continuum sources. As these authors point out, this could cause a severe
underestimate of the Ly$\alpha$ equivalent widths measured on spectra
of high redshift galaxies obtained through a slit.

An analysis of the redshift distribution of the seven confirmed
Ly$\alpha$ objects shows that it is consistent with a random
distribution in the redshift interval selected via the narrow--band
filter.

Despite deep detection limits only S7 and S9 were detected
directly in the combined B and I band images. For S8 and
S10--S13 we registered the broad band images using the
positions of the Ly$\alpha$ sources and determined the 
median image of the five. This procedure allowed the detection 
of broad band emission at the level of B(AB)$\approx$27.3
and I(AB)$\approx$26.8 respectively. This means that S8 and
S10--S13 belong to the faint end of the LF at z=3.
The derived space density of LEGOs is consistent 
with the integrated LF of LBGs down to R=27. However,
only about 20\% of R$<$25.5 LBGs are Ly$\alpha$ emitters
(Steidel et al. 2000).
Hence, either the fraction of Ly$\alpha$ emitters at the faint 
end of the LF is significantly higher than 20\% 
or the space density of galaxies in the field of Q1205-30
is higher than predicted by the LBG LF. The latter would 
indicate that at least some radio quiet QSOs at high redshift 
reside in overdense environments.

It has long been argued that Ly$\alpha$ based searches for high
redshift galaxies were doomed to failure, because even a small amount
of dust would quench the Ly$\alpha$ emission due to the resonant
scattering of the Ly$\alpha$ photons. For this reason it has been
virtually impossible to obtain telescope time for Ly$\alpha$ survey
work. We note here in passing that the recent very succesful Ly$\alpha$
survey by Kudritzki et al. (2000) was aimed at low redshift planetary
nebulae, and the work
that supplied our candidate list was aimed at the imaging of a
quasar absorber. It is virtually certain that neither of those
programmes had been granted telescope time if the aim had been to
search for a sample of Ly$\alpha$ emitters at z=3.

In this paper we find two strong arguments against the presence of
significant amounts of dust in the objects in the faint end of the
high redshift galaxy LF. Firstly we find very large equivalent widths,
in the upper range of theoretical predictions. Secondly we find that
when we calculate the star--formation rate from the continuum flux and
from the Ly$\alpha$ flux independently, we obtain identical results.
Both of those observations are inconsistent with a dust--rich
environment.

The faint continuum magnitudes detected for S8 and S10--S13
prove that Ly$\alpha$ emission is a powerful method by which to probe
the faint end of the galaxy LF at z$\sim$3.
The next step is now to obtain statisitically significant samples,
containing several hundred Ly$\alpha$ selected galaxies,
in order to further constrain the properties of the faint end of the
luminosity function.

\section*{Acknowledgments}
We are grateful for excellent support during our service observations
in January and during the observing run on Paranal in March.
We thank the referee A. Fontana for several comments that clarified our 
manuscript on important points.

\end{document}